\documentclass[12pt]{iopart}

\usepackage{iopams}
\expandafter\let\csname equation*\endcsname\relax

\expandafter\let\csname endequation*\endcsname\relax
\usepackage{physics}
\usepackage[x11names]{xcolor}
\usepackage{booktabs}
\usepackage{graphicx}
\usepackage{color}
\usepackage{enumerate}
\usepackage{cite}
\usepackage[abs]{overpic}
\usepackage[normalem]{ulem}
\usepackage{comment}
\usepackage{etoolbox}
\usepackage[colorlinks=true, linkcolor=blue, urlcolor=blue, citecolor=blue]{hyperref}

\makeatletter
\def\@mkboth#1#2{}
\newlength\appendixwidth
\preto\appendix{\addtocontents{toc}{\protect\patchl@section}}
\newcommand{\patchl@section}{%
  \settowidth{\appendixwidth}{\textbf{Appendix }}%
  \addtolength{\appendixwidth}{1.5em}%
  \patchcmd{\l@section}{1.5em}{\appendixwidth}{}{}%
}
\makeatother

\begin{document}
\title{Wetting and Pattern Formation in Non-Reciprocal Ternary Phase Separation}

\author{Xiao Ma$^{1}$\footnote{Author to whom any correspondence should be addressed: xm245@cam.ac.uk}, Michael E. Cates$^1$}

\address{$^1$DAMTP, Centre for Mathematical Sciences, University of Cambridge, Wilberforce Road, Cambridge CB3 0WA, United Kingdom}

\vspace{10pt}
\begin{indented}
\item[]\today
\end{indented}


\begin{abstract} 

Non-reciprocal interactions are among the simplest mechanisms that drive a physical system out of thermal equilibrium, leading to novel phenomena such as oscillatory pattern formation. In this paper, we introduce a ternary phase separation model, with non-reciprocal interactions between two of the three phases and a spectator phase. Through numerical simulations, we uncover three distinct phase behaviours: a quasi-static regime, characterized by well-defined non-equilibrium contact angles at the three phase contact line; a limit cycle regime, with the three bulk phases rotating around the three phase contact line; and a travelling wave regime, featuring persistent directional motion. We complement our numerical findings with analytical examination of linear stability and the wave propagation speed near equilibrium. Our model provides a minimal framework for extending classical equilibrium wetting theory to active and non-equilibrium systems.

\end{abstract} 
\vspace{2pc}
\maketitle

\section{Introduction}

Wetting, whereby one of two coexisting fluids preferentially comes into contact with a solid boundary, has long been of theoretical, experimental and practical interest. In the past few decades, there have been significant advances in understanding wetting behaviour in thermal equilibrium. In the static case, minimization of the free energy predicts the contact angles formed by a liquid droplet on a surface; this is sometimes interpreted via the Young-Dupre equation, which describes a (tangential) force balance of interfacial tensions \cite{deGennes}. In ternary phase coexistence, such as in three-component emulsions, fluid-fluid-fluid three phase contact lines are instead governed by the equilibrium Neumann law \cite{Neumann, Neumann2}, which sets the contact angles by balancing the interfacial tensions between the three coexisting phases. Closely related extensions to these classical frameworks involve wetting processes on liquid-infused surfaces \cite{LIS} or gel-like deformable substrates \cite{Gel}, which could involve multiple three phase contact lines or elastic stresses within the substrate. 

On the other hand, many wetting phenomena in biological systems occur far from equilibrium, ranging from the spreading of epithelial tissues \cite{epithelial} on a substrate, to the interaction between biomolecular condensates and membrane-bound organelles \cite{wettingbio}. In many of these systems, there is an injection of energy at a microscopic scale, constantly pushing the systems out of equilibrium and challenging established equilibrium principles. In recent years, there have been consistent efforts to understand wetting behaviour in active systems. These include the development of continuum theories for active liquid crystals \cite{ActiveCrystal, ActiveNematic} and active polar liquids \cite{ActivePolar}, where activity enters as an active stress in the Navier-Stokes equations. An alternative approach describes non-equilibrium wetting via an ‘active’ surface, where binary mixtures interact with boundaries through binding and unbinding fluxes that violate detailed balance \cite{ChemActiveWetting}.

A particular set of active systems of interest consists of spherical motile particles: spherical, so that there are no additional order parameters beyond the volume fraction; and motile, so that self-propulsion drives the system out of equilibrium. In these systems, motility-induced phase separation (MIPS) \cite{MIPS} drives particles to accumulate into dense and dilute phases even in the absence of any kind of attractive force; the dense phase in contact with a confining boundary resembles a droplet wetting a wall in the usual equilibrium sense. Crucially, the absence of a free energy in non-equilibrium systems forbids any unique definition of interfacial tensions. One of the most intriguing consequences is the observation of negative interfacial tensions, sparking debates about its physical significance over the stability of active interfaces \cite{NegativeTension2015, NegativeTension2018, NegativeTension2019, NegativeTension2020, NegativeTension2021, NegativeTension2025}. This has prompted exploration of interfacial behaviour and wetting transitions of active particles near solid boundaries with various geometries, including wedge geometry and vertical solid walls, primarily through numerical simulations and experimental realisations \cite{WetNumerical1, WetNumerical2, WetNumerical3, WetNumerical4, Experimental, ActiveTransition}. More recently, there has also been growing interest in quantifying contact angles and generalising the Young-Dupre equation or the Neumann law to its active counterpart, including in self-propelled particle systems \cite{Tangential2020, 2024ActiveYoungJulien, 2025PartialWetting} and in Turing foams \cite{2024FoamAnglesFrey}. As in equilibrium systems, understanding fluid-fluid-fluid contact lines represents a concrete step for exploring interfacial behaviours in wetting.

An alternative to simulating an interacting active particle system is to employ a dynamical field theoretic approach, where coarse-grained density fields evolve via stochastic PDEs that encode the spatiotemporal evolution of the system. At the mesoscopic level, non-equilibrium behaviour manifests through terms that cannot be written as the functional derivative of a free energy; a non-equilibrium field theory can thus be constructed in a top-down manner by explicitly introducing these terms. A celebrated example that describes phase separation in non-equilibrium systems is the non-reciprocal Cahn-Hilliard (NRCH) model \cite{GolestanianPRX, MarchettiNRCH}, which explicitly breaks action-reaction symmetry between two conserved fields. As a minimal model that introduces non-equilibrium features to the lowest order, the NRCH model exhibits a rich spectrum of novel phenomena, including travelling and oscillating patterns \cite{GolestanianPRX, MarchettiNRCH, NRWavesGulati}, arrested coarsening \cite{NRCHArrestedCoarsening}, localised states \cite{NRCHLocalised} and defect solutions \cite{NRCHDefect}. Analytically, oscillations in the NRCH model have been identified to resemble those in the complex Ginzburg-Landau equation \cite{CGLE, GolestanianPRX}. However, the breaking of gauge invariance in NRCH allows for linear oscillations unseen in complex Ginzburg-Landau and has later been shown to emerge from a universal amplitude equation valid near the onset of a conserved-Hopf instability \cite{ThieleAmplitude}, describing large-scale oscillatory instability under conserved dynamics. However, a systematic study of wetting behaviour in models with non-reciprocal interactions remains lacking.

In this paper, we introduce a ternary phase separation model, consisting of two phases with non-reciprocal interactions and a spectator phase. This might be realised experimentally using droplet mixtures with non-reciprocal chemotactic interactions \cite{MotivExp1, MotivExp2}. In the free energy, total symmetry is imposed between the three components that leads to equal interfacial tensions and stable 120-degree contact angles. Under generic conditions, the system relaxes into a configuration with three hexagonal compartments and three-phase contact lines; however, with stripe-like initial configurations, it can also settle to a lamellar pattern without three phase contact lines. These two attractors, hexagonal and lamellar, will be the focus of this study as we depart from equilibrium. 

Weak non-reciprocity breaks the symmetry between the phases, but for both attractors, maintains the stability of patterns in a quasi-static sense. In the hexagonal case, contact angles deviate away from 120 degrees, while in the lamellar case, the pattern begins to travel at a speed that can be captured using perturbative calculations. As non-reciprocity increases, the bulk phases begin to rotate around three phase contact lines, forming limit cycles and ultimately destabilising the contact lines, leading to the emergence of travelling lamellar patterns, which, in this intermediate regime, is itself unstable near the interfaces. At even higher values of non-reciprocity, travelling lamellar patterns can re-stabilise. While our simulations are performed in two dimensions, their interpretation can be extended to three dimensions by assuming homogeneity in the third direction.

The paper is organised as follows. In Section \ref{sec:Model}, we introduce the non-reciprocal ternary phase separation model, and investigate its static behaviour and linear stability analysis. Section \ref{sec:SmallNR} examines the weakly non-equilibrium regime, where both hexagonal and lamellar states remain dynamically stable with slight deviations from their equilibrium counterparts. In Section \ref{sec:LargeNR}, we present numerical observations of pattern formation as higher non-reciprocity moves the system out of the quasi-static regime, including emergence of limit cycles and travelling patterns. We conclude in Section \ref{sec:Conclusion} with discussions and outlook.


\section{A Model for Non-Reciprocal Ternary Phase Separation}\label{sec:Model}

\subsection{The equilibrium model}

We start by considering a Flory-Huggins model that describes symmetric ternary phase separation in thermal equilibrium. The bulk free energy density $f$ of a three-component conserved mixture is expressed through two independent scalar order parameters $\Phi_1$ and $\Phi_2$ \cite{Flory, Huggins},
\begin{equation}\label{eq:FH}
    f(\Phi_1, \Phi_2, \Phi_3) = \sum_{i=1}^3\Phi_i\log\Phi_i + \sum_{i<j}\chi\Phi_i\Phi_j+\chi_0\Phi_1\Phi_2\Phi_3
\end{equation}
with $\Phi_3=1-\Phi_1-\Phi_2$. This can be interpreted as a mixture of two species 1, 2 (which will later become active) with volume fractions $\Phi_{1,2}$ in a solvent of volume fraction $\Phi_3$. The first term in \eqref{eq:FH} represents entropic contributions that favours mixing of the components, while the remaining couplings $\chi$ and $\chi_0$ can be tuned to promote phase separation. We perform a Taylor expansion of the bulk free energy density in terms of deviations $\phi_i$ from the homogeneously mixed state $\Phi_i=1/3+\phi_i$, subject to $\sum\phi_i=0$, to obtain a Cahn-Hilliard-like free energy density \cite{CahnSpinodal, CahnHilliard},

\begin{equation}\label{eq:bulkfree}
    F=\frac{a}{2}\sum_{i=1}^3 \phi_i^2 + \frac{c}{3}\sum_{i=1}^3\phi_i^3 + \frac{b}{4}\sum_{i=1}^3\phi_i^4+\frac{\kappa}{2}\sum_{i=1}^3\left(\nabla\phi_i\right)^2
\end{equation}
where the coefficients are given by $a=3-\chi-\chi_0/3$, $c=\chi_0-9/2$ and $b=9$. The truncation to quartic order is justified for small values of the order parameters $\phi_i$. The derivation of the expression \eqref{eq:bulkfree} is presented in \ref{app:Model}. 

Notably different from the standard Cahn-Hilliard (Model B) theory for binary mixtures, our bulk free energy density contains a cubic term. While this cubic term renders the transition from mixed to phase separation discontinuous, it is essential: without it, the local part of the free energy density reduces to
\begin{equation}\nonumber
    f=\frac{a}{2}\sum_{i=1}^3\phi_i^2+\frac{b}{4}\left(\sum_{i=1}^3\phi_i^2\right)^2 \quad \sum_{i=1}^3\phi_i=0
\end{equation}
resulting in a highly degenerate free energy landscape with a continuous family of minima related by rotations in composition space. The cubic term breaks the symmetry and fixes three minima in composition space, each dominated by a single component. To avoid binary (rather than ternary) phase separations becoming the energy minima, we require $c<0$. Additionally, and as usual \cite{CahnHilliard}, we include gradient terms $\sum\kappa(\nabla\phi_i)^2$ to penalise sharp spatial variations in $\phi_i$. To preserve symmetry between the three phases, the gradient coefficients $\kappa$ have been chosen to be equal. Throughout the simulations in this work, we use $a=-0.1, b=9,c=-1, \kappa=0.4$. In particular, this choice of $\kappa$ ensures that the interface width is in the appropriate range for the numerical study made later of the Neumann angles: much larger $\kappa$ would produce narrow interfaces which are difficult to be resolved accurately on a grid, whereas significantly smaller $\kappa$ would make it harder to locate the interface. It can be set to unity through a rescaling of the length scale and does not change our results qualitatively. 

A linear stability analysis of how a homogeneous phase responds to perturbations in the two independent scalar order parameters ($\phi_1$, $\phi_2$) reveals the spinodal region, within which the system becomes locally unstable to such perturbations and spontaneously phase separate according to conserved dynamics, governed by the equations
\begin{subequations}
\begin{align}
    \partial_t\phi_1&=\nabla^2\frac{\delta\mathcal{F}}{\delta\phi_1}\\
    \partial_t\phi_2&=\nabla^2\frac{\delta\mathcal{F}}{\delta\phi_2}
\end{align}
\end{subequations}
where $\mathcal{F}=\int d\mathbf{r}\;F$ is the free energy and we have set the mobilities for the two species to be equal. While dynamical noise has been introduced in non-reciprocal models and is known to play a crucial role in understanding phase transitions \cite{Vitelli}, it typically does not alter the phase behavior far from criticality \cite{GolestanianPRX, NRCHSarah, Brauns}. In particular, by setting $c$ to be finite, we remain away from the critical transition.

\begin{figure}[t]
    \centering

        \includegraphics[width=0.37\linewidth]{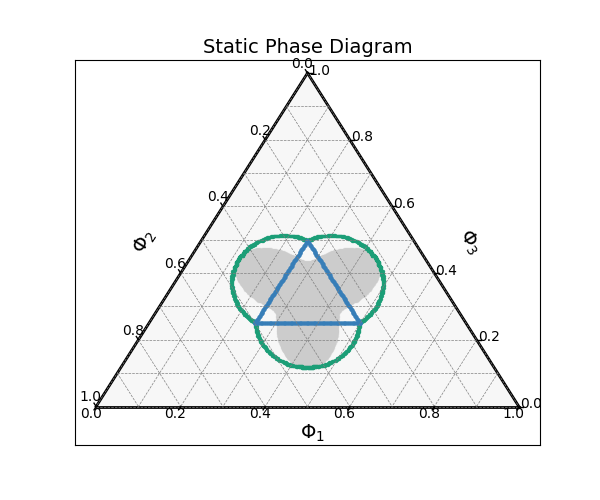}\includegraphics[width=0.31\linewidth]{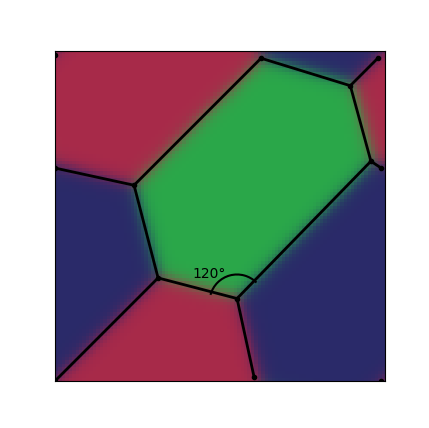}\includegraphics[width=0.31\linewidth]
        {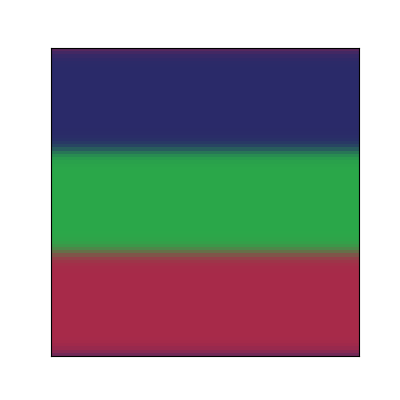}

    \caption{Left panel: Static phase diagram of the symmetric ternary phase separation model. Inside the binodal region (bounded by the green curves), coexistence is favoured over mixing. The intersection points of these curves mark the vertices of the triple phase separation region, inside which the coexistence of the three phases is globally energetically preferred. The spinodal region (shaded area) indicates compositions where spontaneous phase separation occurs. Notably, the spinodal region does not fully cover the triple-phase separation region, since compositions near those of the three coexisting phases, when homogeneous, are trapped in local free energy minima. Middle panel: An example of an equilibrium state of coexistence (under periodic boundary conditions on the square domain), featuring a three-phase coexistence whose contact line is bounded by 120-degree angles. Right panel: An example of an equilibrium lamellar state of coexistence without three phase contact lines. The parameters used throughout the simulations in this paper are $a=-0.1$, $b=9$, $c=-1$ and $\kappa=0.4$.}

    \label{fig:static}
\end{figure}

On the other hand, the binodal region corresponds to uniform states with compositions that favour phase coexistence and are therefore globally unstable. Outside the spinodal and inside the binodal region, the homogeneous system is trapped in a local free energy minimum, and must overcome a free energy barrier in order to reach the phase-separated state that has a lower free energy; this cannot be done without noise-induced nucleation. The spinodal and binodal regions for our equilibrium model \eqref{eq:bulkfree} can be computed analytically (as presented in \ref{app:Model}), and are shown diagrammatically in the left panel of Fig. \ref{fig:static}.

While in principle interfacial profiles between coexisting phases can be analytically obtained by minimising the free energy, this is generally difficult to calculate. Approximations of interfacial free energies are often made under certain simplifying assumptions, such as neglecting the adsorption of the third phase at the interface between the other two \cite{CahnHilliard, Kosmrlj}. Here we do not make use of such assumptions. Instead, we exploit the symmetry between the three phases. While this symmetry does not enable analytical solutions, it ensures that the interfacial tensions between any two coexisting phases, which are well-defined at equilibrium, are identical. Consequently, at any three phase contact line, Young-Dupre equation enforces the equilibrium contact angles to be 120 degrees. An illustration of this using numerical simulations of the system can be found in the middle panel of Fig.~\ref{fig:static}. It is worth noting that the configuration with three hexagonal compartments in the middle panel of Fig.~\ref{fig:static} is the generic attractor for homogeneous initial configurations with equal (or near equal) amounts of the three phases, under periodic boundary conditions. On the other hand, lamellar configurations without a three phase contact line, as illustrated in the right panel of Fig.~\ref{fig:static} are local attractors that do not relax in the absence of noise or activity. In the following sections, we give qualitative descriptions of how both phase attractors are modified in the presence of non-reciprocal interactions.

\subsection{The non-reciprocal model and linear stability analysis}
To describe a non-equilibrium system, we need to include active forcing terms that cannot be derived as functional derivatives from a free energy. To lowest order, this can be achieved by adding an action-reaction symmetry breaking term, as first introduced in the NRCH model \cite{GolestanianPRX}. The modified non-equilibrium dynamics are given by:
\begin{subequations}\label{eq:NREqns}
\begin{align}
    \frac{\partial\phi_1}{\partial t}&=\nabla^2\left[\frac{\delta\mathcal{F}}{\delta\phi_1}+\alpha\phi_2\right]\\
    \frac{\partial\phi_2}{\partial t}&=\nabla^2\left[\frac{\delta\mathcal{F}}{\delta\phi_2}-\alpha\phi_1\right]
\end{align}
\end{subequations}
where the mobility matrix has been set to the identity. While this convenient choice of the mobility matrix breaks the total symmetry between $\phi_{1,2,3}$ at the dynamic level, as can be seen from the anisotropy of the (redundant) $3\times3$ diffusion matrix, this does not affect the physical utility of the model. In particular, our primary interpretation of Eq.~\ref{eq:NREqns} is in terms of two independent scalar fields ($\phi_1$, $\phi_2$) whose dynamics are non-reciprocally coupled, but in whose absence there is a free energy that effectively gives rise to a vapour phase (or empty space) alongside two liquids (respectively, rich in species 1 and species 2). However, Eq.~\ref{eq:NREqns} can also be formulated from the perspective of an incompressible ternary phase separation. This admits an interpretation in which non-reciprocal interactions arise between all pairs of the three species present. Clearly, the chosen interpretation has no effect on the actual evolution of the system and we stick to the two-species picture for clarity below.

Like the NRCH model, this modification introduces a type of `chaser-chased' type dynamics between the order parameters $\phi_1$ and $\phi_2$. Our model is a minimal framework for describing non-reciprocal interactions within a symmetric ternary phase separation, which cannot be captured by the original NRCH model. This non-reciprocal coupling also eliminates any possibility of analytically determining interfacial profiles between coexisting phases. While in non-reciprocal models with simpler cross-couplings between $\phi_1$ and $\phi_2$ (such as NRCH), one can define pseudodensities and a pseudopressure to perform a generalised Maxwell construction \cite{NRCHGreve, NRCHDuan}, the cross-coupling structure in our model prevents a similar approach. Moreover, even when such constructions are possible, these pseudovariables generally lack clear physical interpretations, unlike their equilibrium counterparts.

Importantly, non-reciprocal interactions remove a clear definition for interfacial tensions. While in active field theories such as Active Model B(+), multiple interfacial tensions can be defined to describe the rich phenomenology induced by activity \cite{AMBReview}, an analogous construction is not yet established in non-reciprocal models. Hence, as in the foundational work of Active Model B(+) \cite{NegativeTension2018}, the first line of attack must be primarily numerical to establish the phenomenology.

\begin{figure}[!ht]
    \centering
    \includegraphics[width=\linewidth]{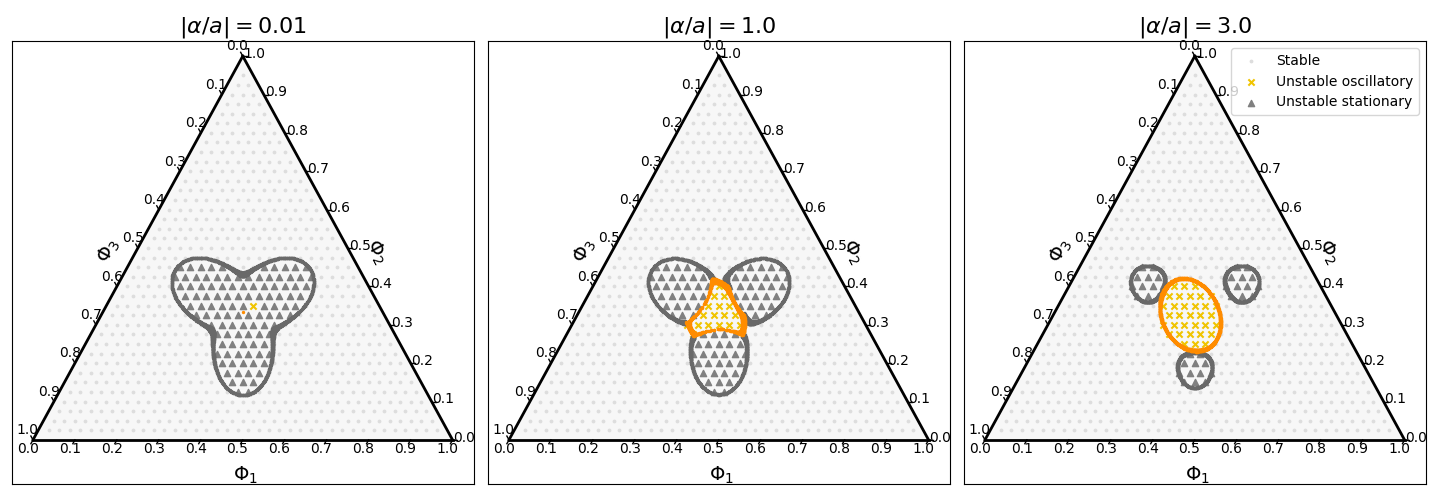}
    \includegraphics[width=\linewidth]{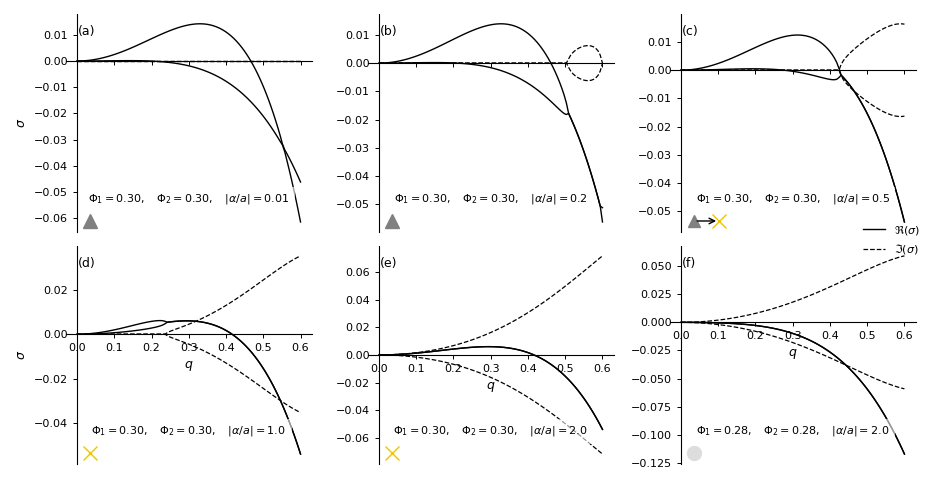}
    \caption{Upper panels: Linear stability analysis of homogeneous compositions at different levels of non-reciprocity, measured by the dimensionless quantity $|\frac{\alpha}{a}|$. Eigenvalue behaviour is categorised into unstable oscillatory (real parts of complex eigenvalues are positive for certain wavenumbers $q$, labeled as orange crosses), unstable stationary (real parts of all complex eigenvalues (if any) are negative, labeled as gray triangles), and purely stable (labeled as light gray dots). Lower panels: Eigenvalue spectra as a function of wavenumber $q$, with solid and dashed lines representing the real and imaginary parts of the eigenvalues respectively. Stability types are indicated in the lower left of each panel using the same markers as in the upper panels. Panels (a)-(e) illustrate an exceptional point transition as non-reciprocity is increased for a fixed composition. Panels (e)-(f) exhibit the conserved Hopf bifurcation induced by varying composition for fixed non-reciprocity.}
    \label{fig:LinStab}
\end{figure}

A linear stability analysis on the non-reciprocal model \eqref{eq:NREqns} reveals the instability structure of a spatially homogeneous state under small spatial perturbations. In the absence of non-reciprocal interactions, the eigenvalues of the linearised stability matrix are purely real: two negative eigenvalues indicate complete stability, whereas the emergence of a positive eigenvalue signals instability toward binary or ternary phase separation. As the level of non-reciprocity $\alpha$ is increased, pairs of complex eigenvalues emerge for a range of modes of finite wavenumber $q$. Initially, these complex eigenvalues have negative real parts, corresponding to damped oscillatory modes that do not manifest pattern formation. The resulting linear dynamics thus qualitatively resemble an equilibrium ternary phase separation, with no significant oscillatory behaviour. At a critical value of $\alpha$ (which may vary depending on the composition of the homogeneous initial state), the real parts of these complex eigenvalues cross zero and become positive, which indicates the appearance of oscillatory behaviour in stationary patterns and is typically described a Hopf bifurcation. In the special case where the imaginary parts of the complex eigenvalues are also zero during this change of stability, it corresponds to a Takens-Bogdanov bifurcation with a higher co-dimension of 2 \cite{Takens1} and is reflected as the orange stability borders in the upper panels of Fig.~\ref{fig:LinStab}. The Takens-Bogdanov bifurcation is also referred to as an exceptional point transition, particularly in models in the presence of non-reciprocal interactions \cite{Vitelli, Brauns}. 

Although linear stability analysis usually only captures the initial growth of perturbations just beyond homogeneous configurations, our numerical simulations reveal that the phase diagram of eigenvalue behaviour provides insights into complex long-time dynamics. At sufficiently small non-reciprocity as shown in the top left panel in Fig.~\ref{fig:LinStab}, most initial configurations evolve into quasi-static long-time behaviour, featuring stable three phase contact lines or slowly travelling, stable patterns. In contrast, larger values of non-reciprocity render most configurations inside the three phase coexistence triangle unstable and oscillatory as demonstrated in the top middle and top right panels in Fig.~\ref{fig:LinStab}, leading to dynamic, and sometimes chaotic, pattern formation in the long time limit. The details of these behaviour will be discussed in the following sections.

Dispersion relations are plotted in the second and third rows of Fig.~\ref{fig:LinStab}. As discussed, increasing non-reciprocity induces an exceptional point transition, where complex eigenvalues acquire positive real parts and oscillatory modes become amplified rather than damped, as exhibited in panels (a)-(e). Varying the homogeneous state composition and/or the non-reciprocity can also induce a conserved Hopf bifurcation (panels (e)-(f)); it is worth noting that our model \eqref{eq:bulkfree}, \eqref{eq:NREqns} is indeed of the general form proposed in \cite{ThieleAmplitude} that gives rise to such large-scale oscillatory transitions between stability and instability.

\section{Quasi-Static behaviour for small non-reciprocal values}\label{sec:SmallNR}

Any finite strength of non-reciprocal interaction will drive the system to depart from thermal equilibrium, where there is no longer an unambiguous notion of free energy or interfacial tensions. Nevertheless, when the non-reciprocity is sufficiently small, the system exhibits quasi-static behaviour that closely resembles its equilibrium counterpart. We characterise this quasi-static regime by focusing on perturbations to the two equilibrium attractors: the configuration with three hexagonal compartments (with three phase contact lines), and the lamellar structure (without three phase contact lines).

\subsection{Contact angles in the quasi-static regime}
In equilibrium, symmetric ternary phase separation features interfaces intersecting at 120 degrees at three phase contact lines, which is a direct consequence of identical interfacial tensions. We initialise the system in a configuration illustrated in the middle panel of Fig.~\ref{fig:static} and numerically analyse the evolution of Neumann angles at different levels of non-reciprocity.

\begin{figure}
    \centering
    \includegraphics[width=\linewidth]{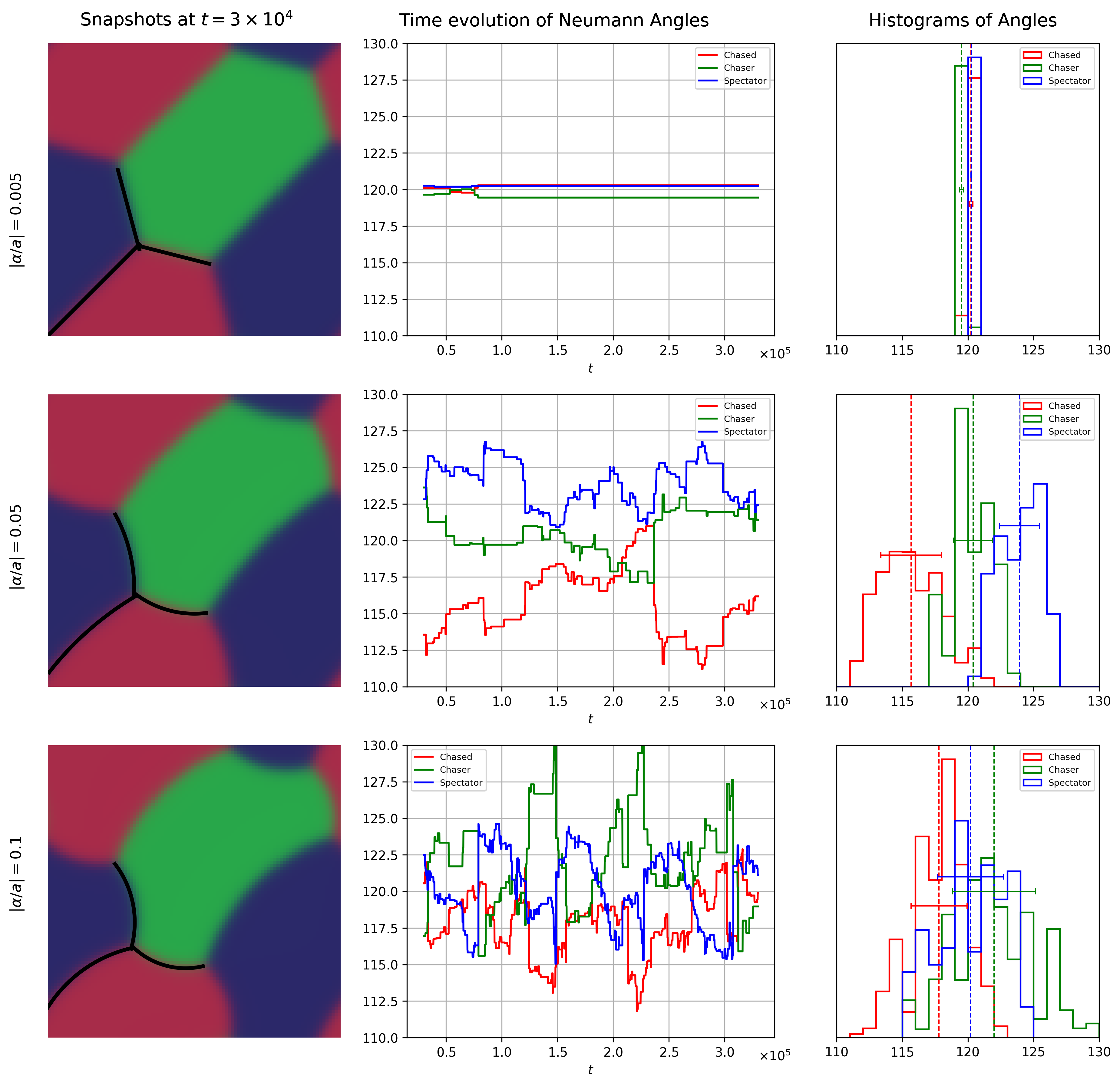}
    \includegraphics[width=\linewidth]{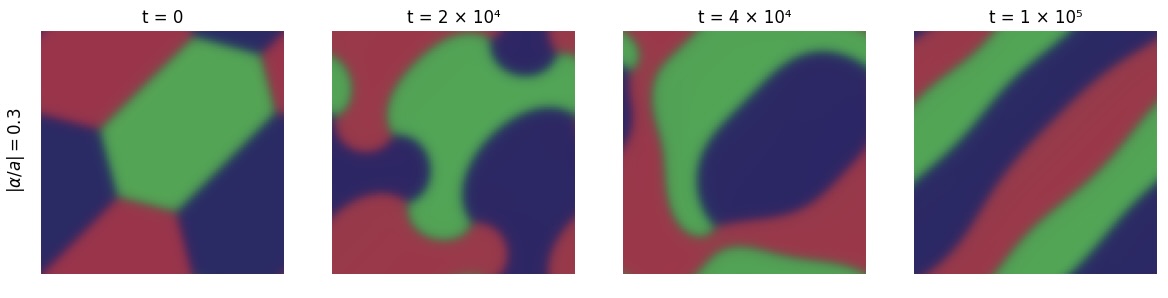}
    \caption{Illustration of quasi-static and non-quasi-static behaviour of Neumann angles. First three rows: snapshots, time evolution, and histograms of Neumann angles at varying levels of non-reciprocity. Snapshots include the linear/circular fits (plotted as black lines/arcs) used to obtain the angles. Time evolutions are initialised in the equilibrium configuration and recorded from $t=3\times 10^4$ after initial transients from the equilibrium configuration. Histograms include the mean and standard deviation of the angle fluctuations. Final row: snapshots of destabilised three phase contact lines joined by three hexagonal compartments. Movies exhibiting quasi-static and chaotic behaviours are provided in the Supplemental Material as ‘Mov 3a’ and ‘Mov 3b’, corresponding to the second row and the final row respectively.}
    \label{fig:Angles}
\end{figure}

Although any finite non-reciprocal interaction precludes a formal definition of interfacial tensions, contact angles remain close to 120 degrees with only small standard deviations, as shown in the first row of Fig.~\ref{fig:Angles}. The configuration with three hexagonal compartments remains stable under slightly stronger non-reciprocity, featuring curved interfaces that can be approximated as circular arcs, as in the snapshot in the second row of Fig.~\ref{fig:Angles}\footnote{Curvature fields are plotted in Fig.~\ref{fig:Curvature} to support this circular arc approximation.}. Crucially, while small errors inherent to the numerical scheme (as discussed in~\ref{app:Numerical}) and minor asynchronous curvature evolution due to limited resolution may both give rise to short-lived measurement fluctuations, the mean values of the Neumann angles exhibit clear deviations of around five degrees away from 120 degrees. This confirms genuine non-equilibrium contact angle behaviour. In addition, non-reciprocity induces a translational motion of this configuration in a direction roughly perpendicular to the longest interface between the chaser (green) phase and the chased (red) phase. While the direction and speed of such translational motion generally depend on the specific local geometry of the configuration, we will give a theoretical prediction for the propagation speed in a simple lamellar pattern in Subsection \ref{subsec:PerturbCalc} below.

Further increase in non-reciprocity invokes non-quasi-static behaviour: contact angles begin to exhibit dynamic fluctuations whose magnitude cannot be attributed to the numerical artefacts discussed above (third row of Fig.~\ref{fig:Angles}); importantly, these genuine fluctuations eventually lead to destabilised three phase contact lines (final row). In this final case, the contact angles visibly deviate far enough from 120 degrees to produce noticeable disruption to the initial structure (with three hexagonal compartments), and the system usually undergoes chaotic motion before settling into oscillatory or travelling patterns, as will be explained in the Section~\ref{sec:LargeNR} below.

\subsection{Perturbative calculation for propagation speed of a periodic lamellar pattern}\label{subsec:PerturbCalc}
In equilibrium, while phase separation in the presence of a three phase contact line is the global free energy minimum, initialising the system with a lamellar pattern traps it in a local minimum preventing relaxation. Small non-reciprocity values do not disrupt this configuration but cause it to become a stable travelling pattern with a finite propagation speed. Closely following the analytical method in \cite{wavespeed, VANSAARLOOS_2003}, we calculate the velocity of travelling patterns near their onset in a one-dimensional periodic domain from $-L$ to $+L$. From Section \ref{sec:Model}, the dynamical equations for the scalar fields are
\begin{align}\label{eq:DynamicalEq}
    \partial_t\begin{pmatrix}\phi_1 \\ \phi_2 \end{pmatrix}=\nabla^2\left[\begin{pmatrix}\frac{\partial f}{\partial\phi_1}\\\frac{\partial f}{\partial\phi_2}\end{pmatrix} + \begin{pmatrix} 0 & \alpha \\ -\alpha & 0 \end{pmatrix} \begin{pmatrix}\phi_1 \\ \phi_2 \end{pmatrix}-\kappa\begin{pmatrix} 2 & 1\\ 1 & 2 \end{pmatrix}\nabla^2\begin{pmatrix}\phi_1 \\ \phi_2 \end{pmatrix}
\right]
\end{align}
Importantly, we make the assumption that the interfacial profiles in the comoving frame of the travelling pattern are only small perturbations away from the equilibrium profiles, i.e.
\begin{equation}\label{eq:perturb}
    \phi_i(\xi=x-vt,t)=\phi_i^{(0)}(\xi,t)+\epsilon\phi_i^{(1)}(\xi,t)+O(\epsilon^2)
\end{equation}
where $\epsilon$ is a small parameter measuring the non-reciprocity, which we set to be of the same order as $\alpha$, or equivalently $\alpha=\bar\alpha\epsilon$, $O(\bar\alpha)=O(1)$. Here, $\phi_i^{(0)}(x)$ denote the equilibrium interfacial profiles with $x$ being the coordinate perpendicular to the interface, and satisfy \eqref{eq:DynamicalEq} to leading order in $\epsilon$
\begin{align}\label{eq:order0}
    \kappa\begin{pmatrix} 2 & 1\\ 1 & 2 \end{pmatrix}\partial_{x}^2\begin{pmatrix}\phi_1^{(0)} \\ \phi_2^{(0)} \end{pmatrix}=\left.\begin{pmatrix}\frac{\partial f}{\partial\phi_1}\\\frac{\partial f}{\partial\phi_2}\end{pmatrix}\right|_{(\phi_1^{(0)},\phi_2^{(0)})}
\end{align}

At $O(\epsilon)$, we now also assume that the velocity is of $O(\epsilon)$ or equivalently $v=\bar{v}\epsilon$ where $\bar v$ is an $O(1) $ quantity. This assumption will be justified self-consistently when matching orders. The $O(\epsilon)$ equations in the comoving frame read

\begin{subequations}\label{eq:MovingFrame}
\begin{align}
    -\bar{v}\frac{d\phi_1^{(0)}}{d\xi}&=\frac{d^2}{d\xi^2}\left[\left.\frac{\partial^2f}{\partial\phi_1^2}\right|_{\{\phi_i^{(0)}\}}\phi_1^{(1)}+\left.\frac{\partial^2f}{\partial\phi_1\partial\phi_2}\right|_{\{\phi_i^{(0)}\}}\phi_2^{(1)}-2\kappa\frac{d^2}{d\xi^2}\phi_1^{(1)}-\kappa\frac{d^2}{d\xi^2}\phi_2^{(1)}+\bar{\alpha}\phi_2^{(0)}\right]\\
    -\bar{v}\frac{d\phi_2^{(0)}}{d\xi}&=\frac{d^2}{d\xi^2}\left[\left.\frac{\partial^2f}{\partial\phi_2^2}\right|_{\{\phi_i^{(0)}\}}\phi_2^{(1)}+\left.\frac{\partial^2f}{\partial\phi_1\partial\phi_2}\right|_{\{\phi_i^{(0)}\}}\phi_1^{(1)}-\kappa\frac{d^2}{d\xi^2}\phi_1^{(1)}-2\kappa\frac{d^2}{d\xi^2}\phi_2^{(1)}-\bar{\alpha}\phi_2^{(0)}\right]
\end{align}
\end{subequations}

These can be integrated twice with respect to $\xi$ to give
\begin{align}\label{eq:SLeq}
    -\bar{v}\begin{pmatrix}\Psi_1^{(0)}\\ \Psi_2^{(0)}\end{pmatrix}=\mathcal{L}\begin{pmatrix}\phi_1^{(1)}\\ \phi_2^{(1)}\end{pmatrix}+\bar{\alpha}\begin{pmatrix}\phi_2^{0}\\ -\phi_1^{(0)}\end{pmatrix} 
\end{align}
with
\begin{equation}
    \Psi_i^{(0)}(\xi):=\int^{\xi}\phi_i^{(0)}(\tilde\xi)d\tilde\xi
\end{equation}
being the indefinite integral (integration constant to be set later) and
\begin{equation}\label{eq:SLOperator}
    \mathcal{L}:=\begin{pmatrix}f_{11}(\xi) & f_{12}(\xi)\\ f_{21}(\xi) & f_{22}(\xi)\end{pmatrix} - \kappa\begin{pmatrix}2 & 1 \\ 1 & 2\end{pmatrix}\frac{d^2}{d\xi^2}
\end{equation}
Here $f_{ij}$ is a shorthand expression for $\partial^2f/\partial\phi_i\partial\phi_j$ evaluated near the equilibrium profiles $(\phi_1^{(0)},\phi_2^{(0)})$. This differential operator $\mathcal{L}$ is a generalised Sturm-Liouville operator (see \ref{app:SL} for definition and proof), which is self-adjoint, i.e.
\begin{equation}
    \langle\begin{pmatrix}\psi_1\\ \psi_2\end{pmatrix},\,\mathcal{L}\begin{pmatrix}\psi_3\\ \psi_4\end{pmatrix}\rangle=\langle\mathcal{L}\begin{pmatrix}\psi_1\\ \psi_2\end{pmatrix},\,\begin{pmatrix}\psi_3\\ \psi_4\end{pmatrix}\rangle
\end{equation}
for any twice continuously differentiable ($C^2$) periodic functions $\psi_{i}$, and inner product defined as
\begin{equation}
    \langle\begin{pmatrix}\psi_1\\ \psi_2\end{pmatrix},\begin{pmatrix}
        \psi_3\\ \psi_4   \end{pmatrix}\rangle=\int_{-L}^{+L}\psi_1\psi_3+\psi_2\psi_4\,dx
\end{equation}
Importantly, note that $\begin{pmatrix}
\frac{d}{d\xi}\phi_1^{(0)}\\
\frac{d}{d\xi}\phi_2^{(0)}
\end{pmatrix}$ is a solution to the homogeneous equation $\mathcal{L}\vec{\phi}$=0 (proof of this intermediate result is presented in~\ref{app:SL}). Therefore, for the Sturm-Liouville differential equation \eqref{eq:SLeq} to be solvable, the Fredholm alternative \cite{PDEbook} requires
\begin{align}\label{eq:solvability}
    \langle \begin{pmatrix}
        \frac{d}{d\xi}\phi_1^{(0)} \\ \frac{d}{d\xi}\phi_2^{(0)}
    \end{pmatrix}, \begin{pmatrix}-\bar{v}\Psi_1^{(0)}-\bar{\alpha}\phi_2^{(0)}\\ -\bar{v}\Psi_2^{(0)}+\bar{\alpha}\phi_1^{(0)}\end{pmatrix}\rangle=0
\end{align}
This solvability condition \eqref{eq:solvability} can also be interpreted through a mechanical analogy, where the parameter $\bar{v}$ has to be fine tuned for a fictitious ‘quasiparticle’ in phase space to have a heteroclinic orbit connecting the two bulk phases, which act as fixed points of the associated dynamical system \cite{wavespeed, Solidification}. \eqref{eq:solvability} produces the value for travelling speed $v$ as proportional to non-reciprocity $\alpha$, dependent on the equilibrium profiles via
\begin{align}\label{eq:SpeedPredict}
    v=\alpha\frac{\left[\int_{-L}^{L}\phi_1^{(0)}\phi_2^{(0)'}-\phi_2^{(0)}\phi_1^{(0)'}d\xi\right]}{\left[\int_{-L}^{L}\Psi_1^{(0)}\phi_1^{(0)'}+\Psi_2^{(0)}\phi_2^{(0)'}d\xi\right]}
\end{align}
In the numerator, we note that the equilibrium reference profiles $\phi_i^{(0)}$ do not obey parity symmetries $\phi_1^{(0)}(x)=-\phi_1^{(0)}(-x),\phi_2^{(0)} (x)=-\phi_2^{(0)}(-x)$ which would otherwise render the numerator zero. Instead, asymmetry in the equilibrium profiles allow the lamellar pattern to set into motion for arbitrarily small non-reciprocity strength $\alpha$. While the functions $\Psi_i^{(0)}$ increase linearly away from the interface, the denominator remains finite. This is because interfacial gradients $\phi_i^{(0)'}$ decay exponentially fast away from the interface; therefore, the denominator integral is contributed primarily at the interface. Note also that there remain degrees of freedom in the integration constants of $\Psi_i^{(0)}$; the arguments for fixing these redundancies are presented in \ref{app:SL}.

\begin{figure}[!ht]
    \centering
    \includegraphics[width=0.35\linewidth]{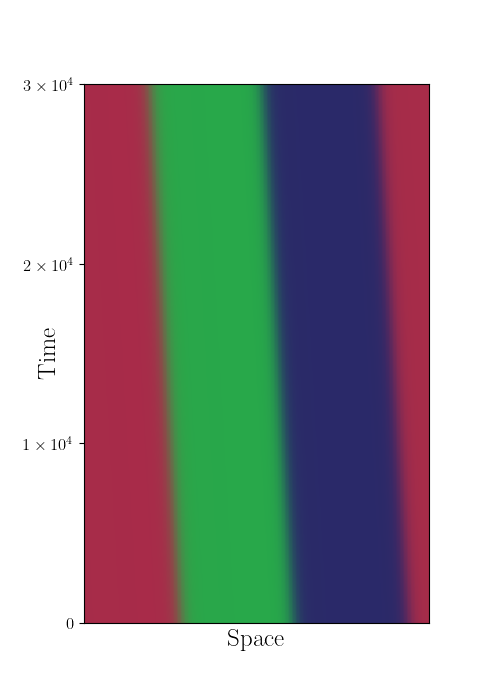}%
    \includegraphics[width=0.55\linewidth]{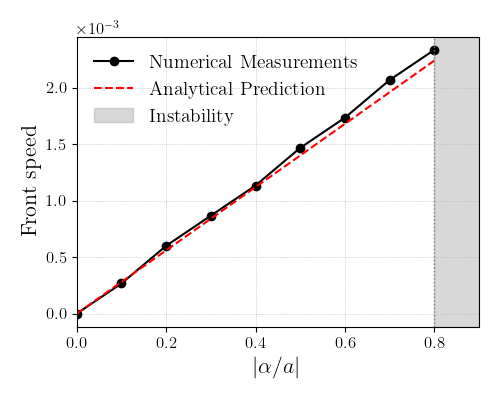}
    \includegraphics[width=0.9\linewidth]{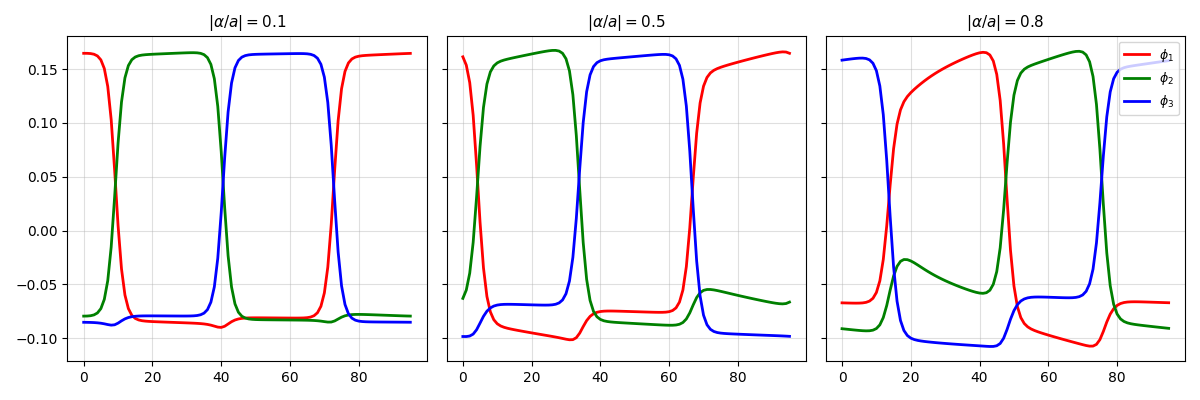}
    \caption{
        Upper left panel: Spatiotemporal plot showing stable propagation of lamellar patterns for relatively small non-reciprocity $|\alpha/a|=0.1$. Upper right panel: Propagation speed analytically predicted \eqref{eq:SpeedPredict} and numerically measured from simulations performed in two dimensions. Lower panel: Concentration profiles perpendicular to the chaser (green) - chased (red) interface at varying levels of non-reciprocity (near-equilibrium, stable propagation, and near onset of instability respectively), obtained from a 1D cross section of the 2D system.
    }
    \label{fig:PerturbSpeed}
\end{figure}

Importantly, the arguments above have implicitly assumed that the perturbation expansion remains valid across the entire domain. This is only true for small non-reciprocity and finite domain sizes so that bulk gradients remain small. In particular, our calculations imply that the bulk must respond instantaneously to movement of the interface, whereas in practice, the diffusive current scales inversely proportional to the wavelength, limiting the validity of \eqref{eq:SpeedPredict} for large non-reciprocity or domain sizes. The one-directional cross sectional profiles, taken perpendicular to the chaser (green) - chased (red) interfaces in the 2D system, are shown in the lower panels of Fig.~\ref{fig:PerturbSpeed} and demonstrate how they deviate further from equilibrium profiles as non-reciprocity increases.

We verify our results by numerically measuring the propagation speed of a two-dimensional lamellar pattern with fixed wavelength, where negligible transverse directions further make the dynamics effectively one-dimensional. As demonstrated in Fig.~\ref{fig:PerturbSpeed}, the perturbation calculations provides an accurate prediction of propagation speed of the lamellar pattern both in the small non-reciprocity regime and well beyond it, up to the point where the travelling pattern ultimately loses stability.  We will examine this loss of stability in the following sections.

As a final note, an ‘interface mode’ method \cite{Brauns} has been shown to give an accurate prediction of the propagation speed in NRCH coupled to non-reciprocal diffusivity. From that perspective, linear unstable modes at the interface grow and saturate at plateau values and are the driving factors for travelling patterns. While in our ternary phase separation model it is difficult to exactly pinpoint the interface, we find that compositions around the interface (e.g. at $\phi_1=\phi_2=0$) also exhibit similar unstable modes. Importantly, whereas in \cite{Brauns} patterns only begin to travel at a finite distance away from equilibrium, here unstable modes can propagate for arbitrarily small $\alpha$ (i.e. $\text{Im}(q^+)>0$, $q^+$ being the marginally unstable wavenumber), matching our prediction using the perturbative method.

\section{Pattern formation far away from equilibrium}\label{sec:LargeNR}
\begin{figure}[!ht]
    \centering
    \includegraphics[width=0.95\linewidth]{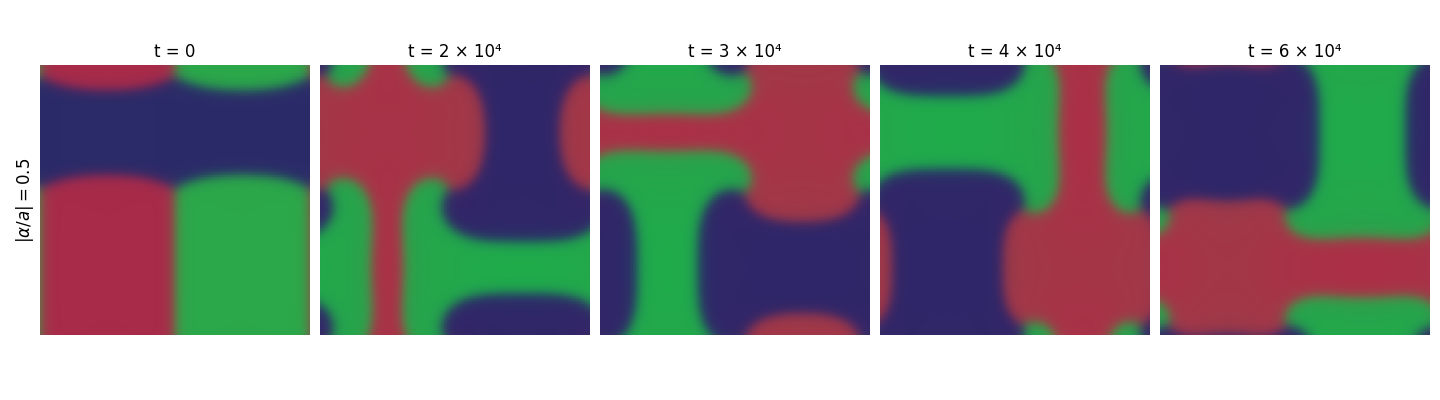}
    \includegraphics[width=0.95\linewidth]{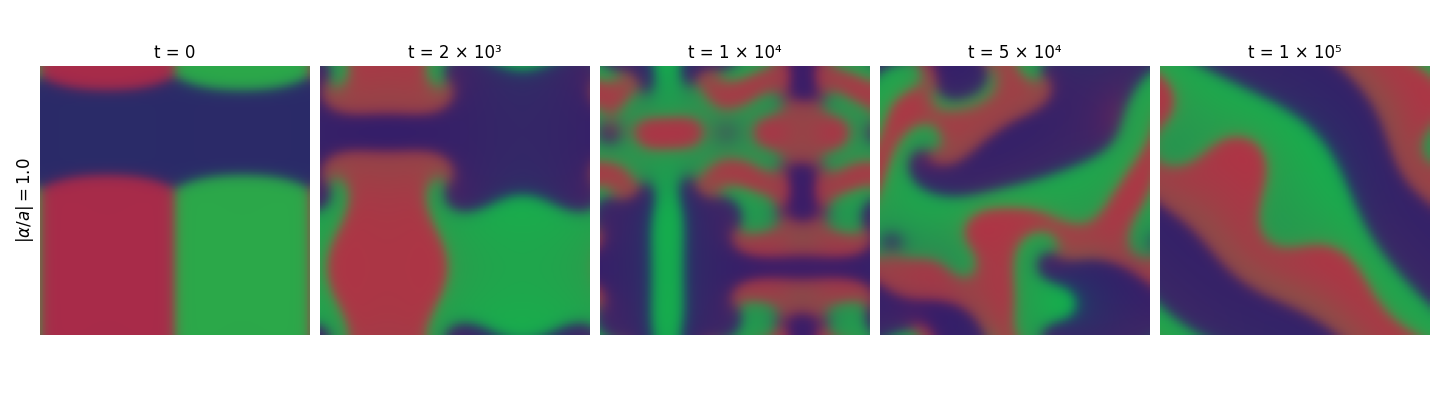}
    \caption{Larger values of non-reciprocity ($|\alpha/a|=0.5$ in first row of the figure) disrupt the quasi-static regime and enable the system to transition into a limit cycle regime. In this figure, time progresses from left to right, with the chaser (green) phase chasing the chased (red) phase around the three phase contact line. Further increase of non-reciprocity ($|\alpha/a|=1$ in second row of the figure) causes three phase contact lines to break, and after undergoing a chaotic transient, the system eventually evolves into a travelling pattern (which itself is unstable at $|\alpha/a|=1.0$, as further illustrated in Fig.~\ref{fig:WaveStab} and Mov 6a). See also corresponding Movies titled ‘Mov 5a’ and ‘Mov 5b’ in the Supplemental Material.}
    \label{fig:LimitCycle}
\end{figure}

Increasing non-reciprocity destabilises the quasi-static regime: the distribution for contact angles shows increased standard deviation, and the system develops oscillatory pattern formation. Just beyond the onset of this oscillatory instability, fully phase separated configurations with three phase contact lines may still persist, but with the phases rotating around the junctions, as demonstrated in Fig.~\ref{fig:LimitCycle}. In this regime, the contact angles fluctuate and no longer settle into well-defined values. The rotational motion around three phase contact lines resembles a rotational defect structure that itself becomes unstable at even higher values of non-reciprocity. A similar loss of stability near spiral defects has also been reported in the non-reciprocal Ising model \cite{NRIsing}. In this high non-reciprocity regime, the rotating configuration breaks down to give travelling lamellar patterns.

\begin{figure}[!ht]
    \centering
    \includegraphics[width=\linewidth]{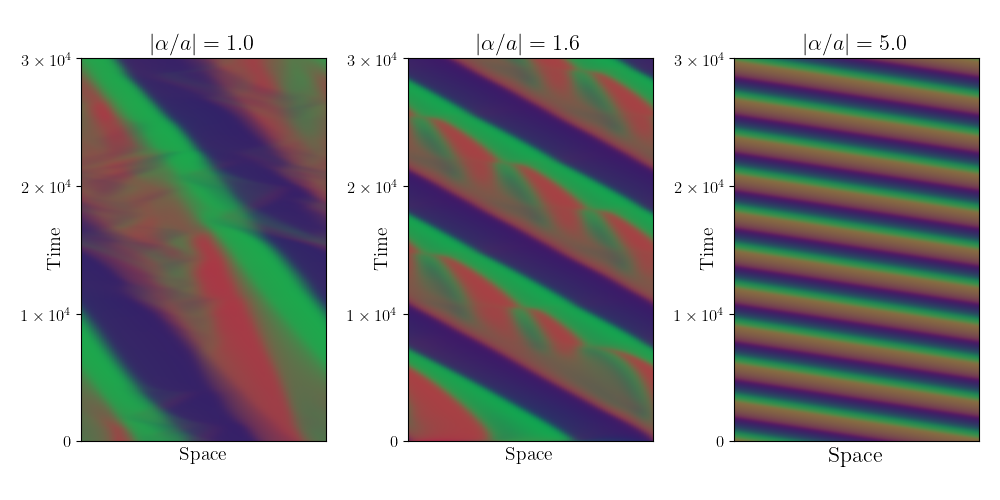}
    \includegraphics[width=\linewidth]{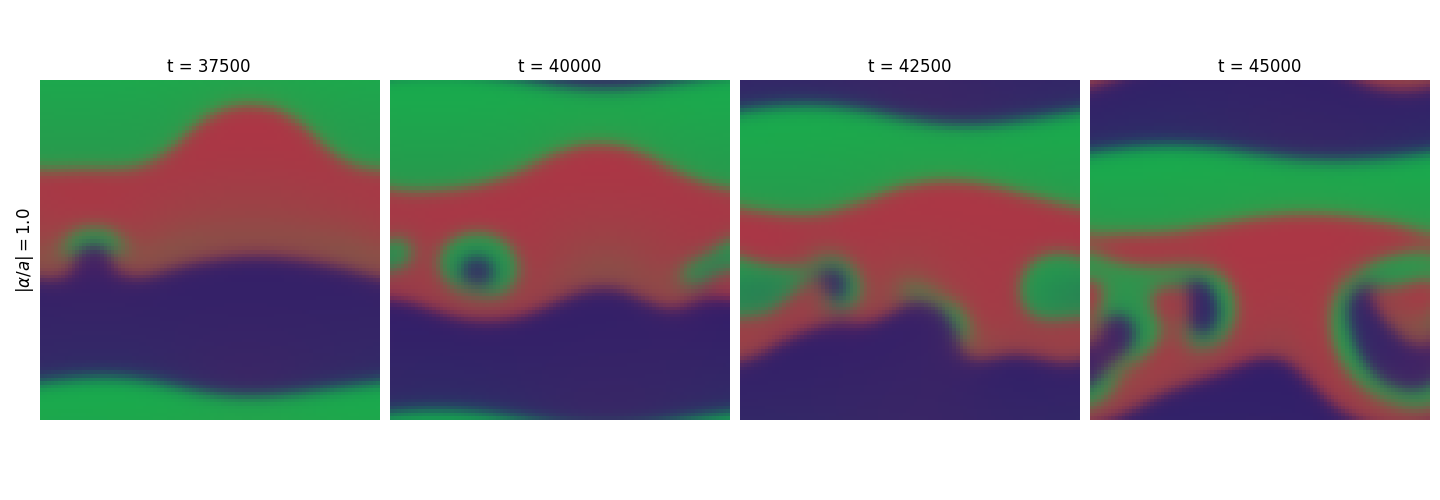}
    \caption{Stability of travelling waves are studied by simulating the dynamics for systems initiated with a fully phase separated lamellar pattern. Upper panels: Spatiotemporal evolution of a 1D slice of the 2D travelling waves at varying levels of non-reciprocity. Stability is lost as the dimensionless non-reciprocity approaches 1, with chaotic behaviour emerging for interfaces involving the chased (red) phase. However, stability of travelling patterns re-emerges in the form of a coherent travelling pattern at higher levels of non-reciprocity (for $|\alpha/a|\gtrsim1.6$). Lower panels: Onset of interfacial stability at $|\alpha/a|=1.0$, caused by nucleation of the chaser (green) phase near the chased (red)-specator (blue) interface. Movies corresponding to chaotic patterns in the left panel (titled ‘Mov 6a’) and coherent travelling patterns in the middle panel (titled ‘Mov 6b’) are provided in the Supplemental Material.}
    \label{fig:WaveStab}
\end{figure}

While carefully prepared initial conditions, such as a fully phase-separated state with three-phase contact lines at intermediate non-reciprocity, can support a limit cycle, more generic setups typically evolve directly into travelling lamellar patterns. As discussed in the Section \ref{sec:SmallNR} above, these travelling patterns are stable, with a propagation speed predictable using perturbative methods. As non-reciprocity $\alpha$ is increased, we observe nucleation of the chaser (green) phase near interface between the chased (red) and the spectator (blue) phase, resulting in instabilities in both chaser-chased (green-red) and chased-spectator (red-blue) interfaces, as demonstrated in Fig.~\ref{fig:WaveStab}. On the contrary, chaser-spectator (green-blue) interface remains stable against perturbations throughout this regime. These instabilities of the non-reciprocal interfaces seem more complex than interference between finite-wavelength modes; instead, they exhibit spatiotemporal chaos. At even larger values of non-reciprocity, the system undergoes a restabilisation; the chaser-chased interface is no longer sharp, and reorganises into a ‘coherent travelling pattern’ (periodic in time but not invariant in the comoving frame), as defined in \cite{VANSAARLOOS_2003}. This stabilisation likely results from the strong propagation (high wave speed $v$), advecting instabilities away before they can fully grow. While the full stability analysis of travelling waves remains to be fully understood, similar instabilities of travelling waves in the form of interfacial undulations and spatiotemporal chaos have also been reported in \cite{Brauns}. 

\section{Conclusion and Outlook}\label{sec:Conclusion}
In this paper, we have introduced, developed, and analysed a minimal model for symmetric ternary phase separation with non-reciprocal interactions between two of the three components. Through a combination of numerical simulations, linear stability analysis, and perturbative approaches, we have uncovered a wide range of dynamic behaviour. Our model captures non-equilibrium dynamics intrinsic to non-reciprocal interactions, including limit cycles, travelling waves and exceptional point transitions. Among our results, most notably we find that contact angles remain well-defined and stable under weak non-reciprocity, and we report the restabilisation of travelling patterns in the form of coherent structures far from equilibrium.

While this work has primarily focused on numerical explorations, it opens several analytical directions for future investigations. For instance, the emergence of well-defined contact angles raises the question of whether an effective interfacial tension (or pseudotension as in Active Model B(+) \cite{AMBReview}) can be meaningfully defined in non-reciprocal systems. Equally intriguing is the mechanism behind the destabilisation and restabilisation of travelling waves; in particular, it would be particularly interesting to understand the transition from chaotic behaviour into a coherent, time-periodic structure. Finally, while we have studied wetting behaviour for a three phase coexistence system, wetting near solid boundaries remains unexplored in this present work. Our choice to in effect replace a solid boundary with a spectator fluid phase may overlook important effects of boundary conditions in active matter systems \cite{ActiveBoundaries}. Extending the current framework to solid substrates could shed light on how physics of wetting transitions or Young-Dupre law are altered by interactions between fluid phases and boundaries, potentially offering new insights into the stability and dynamics of active interfaces.

\section*{Acknowledgements}
We thank Filippo de Luca for useful discussions. XM thanks the Cambridge Commonwealth, European and International Trust, China Scholarship Council, and Trinity College, University of Cambridge for a joint studentship.

\appendix

\section{Derivation of the bulk free energy density and the static phase diagram}\label{app:Model}

The Flory-Huggins free energy density $f$ is expressed via scalar order parameters $\Phi_1, \Phi_2, \Phi_3$ which satisfy the conservation condition $\sum_{i=1}^3\Phi_i=1$,
\begin{equation}
    f(\Phi_1, \Phi_2, \Phi_3) = \sum_{i=1}^3\Phi_i\log\Phi_i + \sum_{i<j}\chi\Phi_i\Phi_j+\chi_0\Phi_1\Phi_2\Phi_3
\end{equation}
Expanding the free energy density around the equally mixed state, $\Phi_i=\phi_i+\frac{1}{3}$ gives

\begin{multline}
    f(\{\phi_i\})=\left[-\log3+\frac{\chi}{3}+\frac{\chi_0}{27}\right]+\left[(-\log3+1+\frac{2\chi}{3}+\frac{\chi_0}{9})\sum_{i=1}^3\phi_i\right]\\
    +\left[\frac{3}{2}\sum_{i=1}^3\phi_i^2+(\chi+\frac{\chi_0}{3})\sum_{i<j}\phi_i\phi_j\right]
    +\left[-\frac{3}{2}\sum_{i=1}^3\phi_i^3+\chi_0\phi_1\phi_2\phi_3\right]+\left[\frac{9}{4}\sum_{i=1}^3\phi_i^4\right]
\end{multline}

where terms have been grouped to indicate their orders in perturbations $\phi_i$. The additive constant can be neglected and the linear term vanishes since $\sum_{i=1}^3\phi_i=0$. We further exploit this conservation condition, in particular
\begin{subequations}
    \begin{align}
        \sum_{i=1}^3\phi_i^2+2\sum_{i<j}\phi_i\phi_j&=\left(\sum_{i=1}^3\phi_i\right)^2=0\\
        \sum_{i=1}^3\phi_i^3-3\phi_1\phi_2\phi_3&=\left(\sum_{i=1}^3\phi_i\right)\left(\sum_{i=1}^3\phi_i^2-\sum_{i<j}\phi_i\phi_j\right)=0
    \end{align}
\end{subequations}

Thus, the bulk free energy density reduces to 
\begin{equation}\begin{split}
    f&=\frac{3-\chi-\chi_0/3}{2}\sum_{i=1}^3 \phi_i^2 + \frac{\chi_0-9/2}{3}\sum_{i=1}^3\phi_i^3 + \frac{9}{4}\sum_{i=1}^3\phi_i^4\\
    &:=\frac{a}{2}\sum_{i=1}^3 \phi_i^2 + \frac{c}{3}\sum_{i=1}^3\phi_i^3 + \frac{b}{4}\sum_{i=1}^3\phi_i^4
\end{split}\end{equation}
confirming \eqref{eq:bulkfree} of the main text.

To determine the coexistent bulk phase compositions of the static system, we minimise \eqref{eq:bulkfree} subject to the constraint $\sum_{i=1}^3\phi_i=0$. This can be  achieved either by introducing a Lagrange multiplier or by rewriting \eqref{eq:bulkfree} as a function of the two independent scalar order parameters $\phi_1$ and $\phi_2$. We adopt the latter method, and requiring $\partial f/\partial\phi_1=\partial f/\partial\phi_2=0$ produces the three energy minima, occupying symmetric positions within the triangular phase diagram of Fig. \ref{fig:static}. Due to total symmetry of the free energy between the three phases, the bulk compositions take the form $(2\phi_0,-\phi_0, -\phi_0)$, $(-\phi_0,2\phi_0, -\phi_0)$, and $(-\phi_0,-\phi_0, 2\phi_0)$, where $\phi_0$ is the positive root to the equation
\begin{equation}
    a+c\phi_0+3b\phi_0^2=0
\end{equation}
Note that this set of solutions is guaranteed as long as $a<0$, $c<0$ and $b>0$, consistent with the parameter choices in this work. 

The binodal lines of the system are found by a common tangent construction, requiring the chemical potentials of each component, $\mu_i=\frac{\partial f}{\partial\phi_i}$, and the pressure, $P=\sum_{i=1}^3\phi_i\mu_i-f$, are equal across coexisting bulk compositions. Due to symmetry between the three phases, the pressure is automatically equal across phases. As for the equal chemical potentials, we consider two symmetric coexisting compositions ($\phi_1$, $\phi_2$, $-\phi_1-\phi_2$), labeled with a superscript (1), and $(\phi_2,\phi_1,-\phi_1-\phi_2)$, labeled with a superscript (2), with $\phi_1\neq\phi_2$ 
\begin{equation}
    \mu_1^{(1)}=\mu_1^{(2)}\Rightarrow\left.\frac{\partial f}{\partial\phi_1}\right|_{(\phi_1,\phi_2,1-\phi_1-\phi_2)}=\left.\frac{\partial f}{\partial\phi_1}\right|_{(\phi_2,\phi_1, 1-\phi_1-\phi_2)}
\end{equation}
which simplifies to
\begin{multline}
    (2\phi_1+\phi_2)\left[a-c\phi_2+b(\phi_1^2+\phi_1\phi_2+\phi_2^2)\right]\\=(\phi_1+2\phi_2)\left[a-c\phi_1+b(\phi_1^2+\phi_1\phi_2+\phi_2^2)\right]
\end{multline}
or equivalently
\begin{equation}
    a+c(\phi_1+\phi_2)+b(\phi_1^2+\phi_1\phi_2+\phi_2^2)=0
\end{equation}
By symmetrising this relation under all permutations of the three scalar order parameters, we recover the binodal lines in the composition space, as shown in Fig.~\ref{fig:static}.

The spinodal region is determined using a linear stability analysis of a perturbed homogeneous state $(\bar\phi_1,\bar\phi_2)$, with
\begin{align}
    \begin{pmatrix}
        \dot\phi_1\\ \dot\phi_2
    \end{pmatrix}
    =-q^2\begin{pmatrix}
        J_{11} & J_{12}\\ J_{21} & J_{22}
    \end{pmatrix}
    \begin{pmatrix}
        \phi_1 \\ \phi_2
    \end{pmatrix}
\end{align}
for perturbations with wavenumber $q$ in the Fourier space. The entries of the Jacobian matrix are given by
\begin{subequations}\label{eq:Jacobian}
\begin{align}
J_{11} &= 2\left(a - c\phi_2 + b(\phi_1^2 + \phi_1\phi_2 + \phi_2^2)\right)+ (2\phi_1 + \phi_2)(2b\phi_1 + b\phi_2), \\
J_{12} &= \left(a - c\phi_2 + b(\phi_1^2 + \phi_1\phi_2 + \phi_2^2)\right)+ (2\phi_1 + \phi_2)(-c + b\phi_1 + 2b\phi_2) + \alpha, \\
J_{21} &= \left(a - c\phi_1 + b(\phi_1^2 + \phi_1\phi_2 + \phi_2^2)\right) + (2\phi_2 + \phi_1)(-c + b\phi_2 + 2b\phi_1) - \alpha, \\
J_{22} &= 2\left(a - c\phi_1 + b(\phi_1^2 + \phi_1\phi_2 + \phi_2^2)\right) + (2\phi_2 + \phi_1)(2b\phi_2 + b\phi_1).
\end{align}
\end{subequations}
Stability of each homogeneous state is then identified by looking at the eigenvalues of the Jacobian matrix \eqref{eq:Jacobian} multiplied by $-q^2$, and expressed diagrammatically in the left panel of Fig.~\ref{fig:static} for the equilibrium case in the main text.

\section{Numerical scheme}\label{app:Numerical}
We perform numerical simulations on a two-dimensional, $96$-by-$96$ square domain with periodic boundary conditions. Space and time are discretised as $\Delta x = \Delta y = 1$ and $\Delta t = 0.01$. All simulations use a finite-difference scheme to integrate the dynamical equations \eqref{eq:NREqns} forward in time. Model parameters are fixed at $a = -0.1$, $b = 9$, $c = -1$, and $\kappa = 0.4$. These values are representative and do not qualitatively alter the results presented. The simulation results are visualised using a RGB colormap, with red (chased) and green (chaser) phases rescaled to enhance contrast.

\begin{figure}[!ht]
    \centering
    \includegraphics[width=0.9\linewidth]{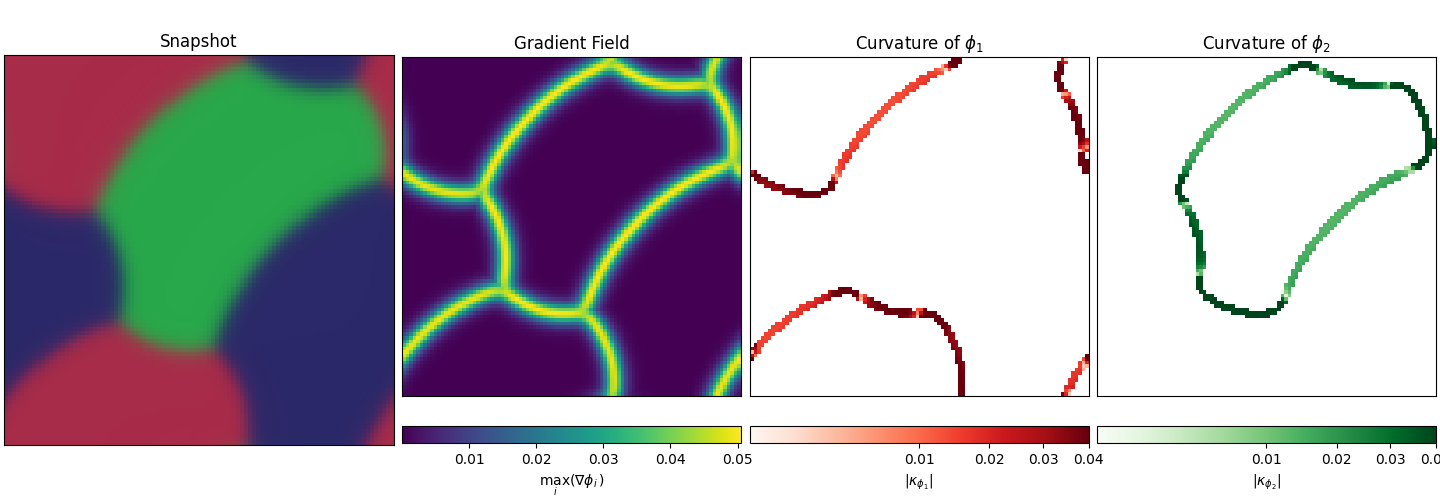}
    \caption{Visualisation of gradient field and curvature fields for a snapshot of the system at $|\alpha/a|=0.1$. From left to right: first panel is a snapshot of the system in our RGB visualisation. Second panel: Maximum gradient magnitude across the two scalar fields, indicating positions of interfaces. Third and fourth panels: Curvature of the two scalar fields, given by $K=\nabla\cdot\frac{\nabla\phi}{|\nabla\phi|}=\frac{\phi_x^2\phi_{yy}+\phi_y^2\phi_{xx}-2\phi_x\phi_y\phi_{xy}}{|\nabla\phi|^3}$. Interfaces typically exhibit near-uniform curvature except in the vicinity of three phase contact lines. However, these curvature magnitudes depend on local geometries: the same type of interface does not necessarily always have the same curvature.}
    \label{fig:Curvature}
\end{figure}

To numerically measure the contact angles, we identify interfaces by thresholding the magnitude of scalar field gradients to construct a binary mask. The resulting structures are skeletonized using the sknw package, which extracts a graph-based representation of the interface. Three phase contact lines are detected as nodes with three neighbours, and each connecting interface is fitted either with a straight line or a circular arc. Close to equilibrium, the linear fit is appropriate for configurations with equal compositions, which is the setup we use. Non-reciprocal interactions introduce curvature that is typically uniform along a given interface, as demonstrated in Fig.~\ref{fig:Curvature}, making circular fits more suitable. Fitting the interfaces enables the calculation of tangents near the three-phase contact lines, from which the contact angles are determined.

There are two potential sources of numerical error in this scheme. First, the choice of a gradient threshold leads to a binary graph with finite interface width of around one pixel, limiting spatial resolution. Subsequent skeletonisation reduces the 2-dimensional binary mask to 1-dimensional graphs, which may distort the geometry of the exact interface, especially near three phase contact lines. The magnitude of this error is around one degree, as exemplified in the near-equilibrium case (first row of Fig.~\ref{fig:Angles}). A second source of error comes from interface fitting. While linear approximations work well for near-equilibrium configurations with symmetric compositions (e.g., configurations with three hexagonal compartments that we study), non-reciprocal interactions create curvature that is only approximately uniform, and may not be fully captured by a circular fit. Nevertheless, based on the curvature measurements shown in Fig.~\ref{fig:Curvature}, we expect the effect of this error to be similarly minor. 

\section{Generalised Sturm-Liouville operators}\label{app:SL}

A Sturm-Liouville operator can be interpreted as a self-adjoint operator under the inner product defined on real-valued functions (with weight function of the inner product $w(x)=1$)
\begin{equation}\label{eq:InnerProduct}
\langle\psi_1(\xi),\psi_2(\xi)\rangle=\int_{-L}^{+L}\psi_1\psi_2\,d\xi
\end{equation}
An operator $\mathcal{L}$ is Sturm-Liouville under \eqref{eq:InnerProduct} if
\begin{equation}
\langle\mathcal{L}\psi_1,\psi_2\rangle=\langle\psi_1,\mathcal{L}\psi_2\rangle
\end{equation}
for all twice continuously differentiable functions $\psi_i(\xi)$ in the domain of $\mathcal{L}$, whose periodic boundary conditions eliminate boundary terms. In this Appendix, we show that $\mathcal{L}$ is a matrix Sturm-Liouville (self-adjoint) differential operator defined on vector-valued functions, via the inner product
\begin{equation}\label{eq:InnerProductVec}
    \langle\begin{pmatrix}\psi_1\\ \psi_2\end{pmatrix},\begin{pmatrix}
        \psi_3\\ \psi_4   \end{pmatrix}\rangle=\int_{-L}^{+L}\psi_1\psi_3+\psi_2\psi_4\,d\xi
\end{equation}

The operator in \eqref{eq:SLOperator} reads
\begin{equation}
    \mathcal{L}:=\begin{pmatrix}f_{11}(\xi) & f_{12}(\xi)\\ f_{21}(\xi) & f_{22}(\xi)\end{pmatrix} - \kappa\begin{pmatrix}2 & 1 \\ 1 & 2\end{pmatrix}\frac{d^2}{d\xi^2}
\end{equation}
which is self-adjoint: the first term is a symmetric Hessian matrix, and the second term is self-adjoint as follows from integration by parts with vanishing boundary terms.

Next, we prove the intermediate result just above \eqref{eq:solvability}. We know that the equilibrium profiles $(\phi_1^{(0)}(\xi),\phi_2^{(0)}(\xi))$ satisfy the equation
\begin{align}
    \left.\begin{pmatrix}\frac{\partial f}{\partial\phi_1}\\\frac{\partial f}{\partial\phi_2}\end{pmatrix}\right|_{(\phi_1^{(0)},\phi_2^{(0)})}-\kappa\begin{pmatrix} 2 & 1\\ 1 & 2 \end{pmatrix}\partial_{\xi}^2\begin{pmatrix}\phi_1^{(0)} \\ \phi_2^{(0)} \end{pmatrix}=0
\end{align}
Differentiating this with respect to $\xi$ produces

\begin{align}
    \begin{pmatrix}f_{11} & f_{12} \\ f_{21} & f_{22}\end{pmatrix}\begin{pmatrix}
        d\phi_1^{(0)}/d\xi \\ d\phi_2^{(0)} /d\xi
    \end{pmatrix}-\kappa\begin{pmatrix} 2 & 1\\ 1 & 2 \end{pmatrix}\partial_{\xi}^2\begin{pmatrix}d\phi_1^{(0)}/d\xi \\ d\phi_2^{(0)} /d\xi\end{pmatrix}=0
\end{align}
where chain rule is applied to the first term. This proves our claim that $\begin{pmatrix}d\phi_1^{(0)}/d\xi \\ d\phi_2^{(0)} /d\xi\end{pmatrix}$ is a solution to the homogeneous equation $\mathcal{L}\vec{\phi}=0$

Given the homogeneous solution $\begin{pmatrix}d\phi_1^{(0)}/d\xi \\ d\phi_2^{(0)} /d\xi\end{pmatrix}$, there exists a redundancy of the perturbative solution, namely if $\phi_i^{(1)}$ is the leading order perturbation and solves \eqref{eq:SLeq}, so is $\phi_i^{(1)}+\lambda\phi_i^{(0)'}$ for arbitrary constant $\lambda$. Indeed, the redundant degree of freedom can be interpreted as a translation of $\lambda\epsilon$ to the equilibrium profile, since
\begin{equation}
    \phi_i(\xi)=\phi_i^{(0)}(\xi)+\epsilon(\phi_i^{(1)}(\xi)+\lambda\phi_i^{(0)'}(\xi))+O(\epsilon^2)=\phi_i^{(0)}(\xi+\lambda\epsilon)+\epsilon\phi_i^{(1)}(\xi)+O(\epsilon^2)
\end{equation}
To eliminate the redundancy and to ensure that the perturbative correction $\phi_i^{(1)}$ does not contain an arbitrary translation of the equilibrium profile, an additional condition is required to ‘fix’ the position of equilibrium profile. This can be done by enforcing that the centre of the interface (of maximum gradient) is fixed in space. Algebraically we let $\left.\mathcal{L}\vec{\phi_i}^{(1)}\right|_{\xi=\xi_{\textbf{centre}}}\equiv 0$. This fixes the integration constant in $\Psi_i^{(0)}$ and gives a unique value for $\bar{v}$.

\section*{References}

\bibliographystyle{unsrt}
\bibliography{ref}

\end{document}